\documentclass[prc, twocolumn,nofootinbib, showpacs]{revtex4} 
\usepackage{graphicx}
\usepackage[sort&compress]{natbib}
\usepackage{subfigure}
\usepackage{amsmath}
\usepackage{amssymb}
\usepackage{amsfonts}
\usepackage{cancel}
\usepackage{mathtools}
\usepackage{color}
\usepackage{hyperref}
\usepackage{ulem}
\usepackage{ulem} 
\begin{document}

\arraycolsep1.5pt

\newcommand{\Ima}{\textrm{Im}}
\newcommand{\Rea}{\textrm{Re}}
\newcommand{\mev}{\textrm{ MeV}}
\newcommand{\be}{\begin{equation}}
\newcommand{\ee}{\end{equation}}
\newcommand{\ba}{\begin{eqnarray}}
\newcommand{\ea}{\end{eqnarray}}
\newcommand{\gev}{\textrm{ GeV}}
\newcommand{\nn}{{\nonumber}}
\newcommand{\cm}{\textcolor{blue}}
\newcommand{\cor}[2]{{\color{red} \sout{#1}} {\bf\color{blue} #2}}

\newcommand{\dtres}{d^{\hspace{0.1mm} 3}\hspace{-0.5mm}}
\def\pmat#1{\begin{pmatrix}#1\end{pmatrix}}

\newcommand{\gls}[1]{{{\color{blue} #1}}}

\def\del{\partial}

\title{Comments on the dispersion relation method to vector-vector interaction}

\author{R. Molina}
\affiliation{
Universidad Complutense de Madrid, Facultad de F\'isicas. Departamento de F\'isica Te\'orica II.  Plaza Ciencias, 1, 28040, Madrid. \& Institute of Physics of the University of Sao Paulo, Rua do Mat\~ao, 1371 -Butant\~a, S\~ao Paulo -SP, 05508-090}
\author{L. S. Geng}
\affiliation{ School of Physics and
Nuclear Energy Engineering \& International Research Center for Nuclei and Particles in the Cosmos \& Beijing Key Laboratory of Advanced Nuclear Materials and Physics, Beihang University, Beijing 100191, China}
\author{E. Oset}
\affiliation{Departamento de F\'{\i}sica Te\'orica and IFIC, Centro Mixto Universidad
de Valencia-CSIC,
Institutos de Investigaci\'on de Paterna, Aptdo. 22085, 46071 Valencia,
Spain}

\date{\today}

 \begin{abstract}
  We study in detail the method proposed recently to study  the vector-vector interaction using the $N/D$ method  and dispersion relations, which concludes that, while for $J=0$, one finds bound states, in the case of $J=2$, where the interaction is also attractive and much stronger, no bound state is found. In that work,  approximations are done for $N$ and $D$ and a subtracted dispersion relation for $D$ is used, with subtractions made up to a polynomial of second degree in $s-s_\mathrm{th}$, matching the expression to $1-VG$ at threshold. We study this in detail for the $\rho - \rho$ interaction and to see the convergence of the method we make an extra subtraction matching $1-VG$ at threshold up to $(s-s_\mathrm{th})^3$. We show that the method cannot be used to extrapolate the results down to 1270 MeV where the  $f_2(1270)$ resonance appears, due to the artificial singularity stemming from the ``on shell" factorization of the $\rho$ exchange potential. In addition, we explore the same method but folding this interaction with the mass distribution of the $\rho$, and we show that the singularity disappears and the method allows one to extrapolate to low energies, where both the $(s-s_\mathrm{th})^2$ and $(s-s_\mathrm{th})^3$ expansions lead to a zero of $\mathrm{Re}\,D(s)$, at about the same energy where a realistic approach produces a bound state.
Even then, the method generates a large $\mathrm{Im}\,D(s)$ that we discuss is unphysical.
\end{abstract}

\maketitle

\section{Introduction}
\label{Intro}
\begin{figure}
  \centering
  \includegraphics[width=0.4\textwidth]{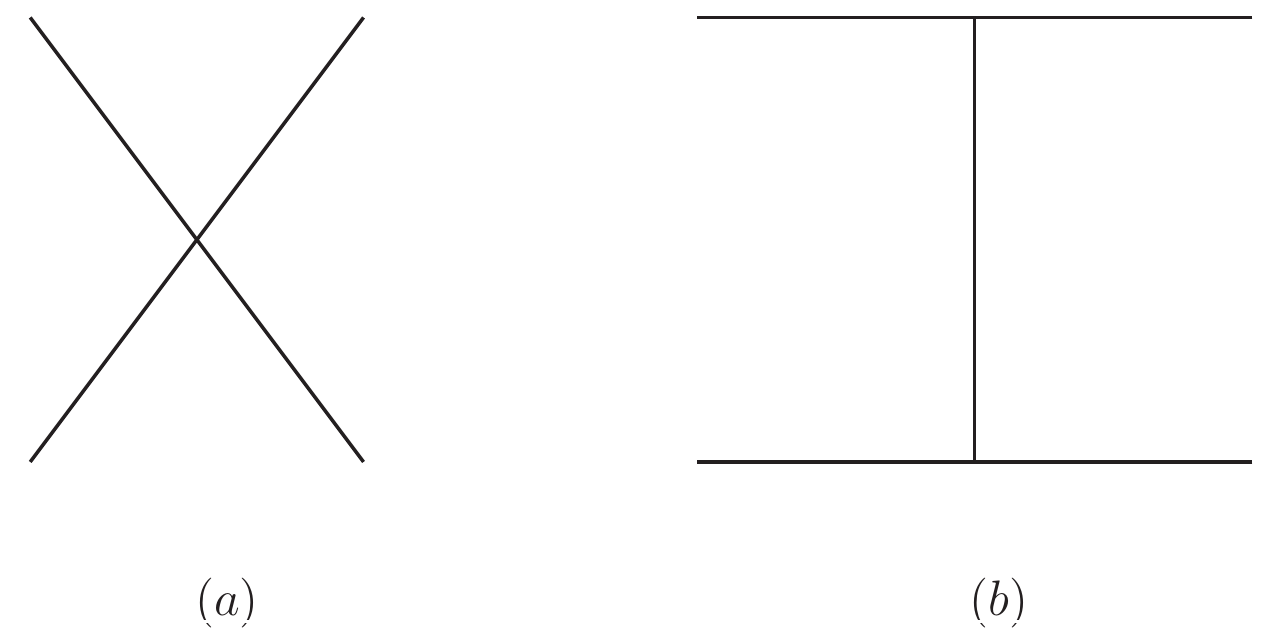} 
  \caption{Terms in the $\rho\rho$ interaction: (a) contact term; (b) $\rho$ exchange term.}\label{fig:1}
\end{figure}

In Ref. \cite{raquel}, the chiral unitary approach for pseudoscalar mesons was extended to the interaction of vector mesons, concretely the $\rho\rho$ interaction,
 using the Bethe Salpeter equation,
 \be
 T=\frac{V}{1-V\,G}\ ,
 \label{eq:bethe}
 \ee
where $G$ is the loop function of the two $\rho$ meson propagator and $V$ the potential, obtained from the local hidden gauge
Lagrangians \cite{hidden1,hidden2,hidden4}, which contains a contact term and the $\rho$ exchange term as shown in Fig.~\ref{fig:1}.

The potential $V$ corresponding to these diagrams for $J=2$ is 
\ba
&&V=V_c+V_{\mathrm{ex}}\nonumber\\
&&V_c=-4\,g^2;\qquad V_{ex}=-8\,g^2 \frac{3\,s-4\,m^2_\rho}{4\,m^2_\rho}\ ,\label{eq:ex}
\ea
where $g=m_\rho/(2\,f_\pi)$ ($f_\pi=93$ MeV). In Ref. \cite{raquel}, an approximation was made, where in the exchange of the $\rho$ meson, the 
$q^2$ term in the propagator of the exchanged $\rho$, $[q^2-m^2_\rho]^{-1}$, was dropped. This is actually what is done to establish the 
link between the local hidden gauge approach, with the exchange of vector mesons, and the chiral Lagrangians. The latter are obtained
from the former neglecting $q^2$ in the propagator of the exchanged vector mesons. 

Two dynamically generated resonances were 
found in isospin $I=0$, one with total angular momentum $J=0$, which could be related to the $f_0(1370)$, and the other with
$J=2$, which was associated to the $f_2(1270)$. The approach was generalized to SU(3) in Ref. \cite{geng}, where other resonances like the $f_0(1710)$ and the 
$f_2'(1525)$ were also obtained.

In Ref. \cite{Gulmez:2016scm}, the method used in Ref. \cite{raquel} was questioned in base to an improved relativistic vertex and keeping the $q^2$ dependence of the exchanged $\rho$ propagator. Eq. (\ref{eq:bethe})
was used in the on-shell factorization of the potential taking the $\rho$-exchange term with the external legs on-shell ($p^2_i=m^2_\rho$). However, the method
developed pathologies since the factorized on-shell $\rho$-exchange term has singularities below threshold,
 giving rise to an unphysical infinity in the potential, and an imaginary part which has also a discontinuity. The method was discussed in Ref. \cite{Geng:2016pmf} and it was shown to provide similar results to Ref. \cite{raquel} close to threshold,
  but to be unsuited for the study of more bound states, as the $f_2(1270)$, because the unphysical singularity of the on-shell potential appeared around the energy of that state. In fact, the conclusion of Ref. \cite{Gulmez:2016scm} was that the $f_2(1270)$ was not obtained in that
  approach and was ruled out as a dynamically generated state from the $\rho\rho$ interaction. The conclusion is surprising because the $f_0(1370)$ appears bound in Refs. \cite{raquel} and \cite{Gulmez:2016scm}, and the potential for $J=2$ is attractive and even more than twice larger than for $J=0$ in the whole 
  relevant energy range. According to basic rules of Quantum Mechanics if we find a bound state with a given potential, another potential with the same range and bigger attractive strength gives rise to a state which is more bound. 
  
  Concerning the total angular momentum of the two states, we should note that while the general rule in Quantum Mechanics is that states with higher L are less bound, because of the centrifugal potential in spherical coordinates, in the present case we have $J=2$ and $J=0$, but both with s-wave, coming from a different combination of spins of the two vectors, and it is the peculiar dynamics of the meson exchange that makes the $J=2$ case more bound. 
  
  The effective range approach, which can give different results for $J=0$ and $J=2$, was invoked as  a possible explanation for this feature in Ref. \cite{Gulmez:2016scm} and more recently in Ref. \cite{Du:2018gyn}. Yet, this argument cannot invalidate the Quantum Mechanics rule. Indeed, if the effective range formula fails to give a bound 
  state in the case of the more bound potential, the only conclusion that one can draw is that the effective range formula,
  \be
  T\sim \frac{-8\pi\sqrt{s}}{\frac{-1}{a}+\frac{1}{2}r_0 p^2-i\,p}\ ,
  \ee
  cannot be extrapolated to the low energies where the bound state will appear.
  
  In Ref. \cite{Geng:2016pmf} it was shown that the singularity and imaginary parts that appear implicitly in the loops of the Bethe-Salpeter equation when the on-shell factorization is done were artificial, because the loop, 
  evaluated exactly in Ref. \cite{Geng:2016pmf}, 
  did not develop any singularity nor had imaginary part
  below threshold. 
  Instead, a method was proposed that kept the $q^2$ dependence of the $\rho$ exchanged propagator in the loops and gave rise unavoidably to a bound state both in $J=0$ and $J=2$.

  To the end of Ref. \cite{Gulmez:2016scm} a different method was proposed based on the N/D method, however, solved perturbatively. This method has been recently used in Ref. \cite{Du:2018gyn} and extended to SU(3) to match with the results obtained in Ref.~\cite{geng}, with the conclusion that 
  while the method provides very similar results to \cite{geng} for small binding energies, it does not provide bound states in $J=2$, as the $f_2(1270)$ and $f'_2(1525)$. 
  The purpose of the present paper is to show in detail why and how this perturbative N/D method fails when one goes to large binding energies. Actually the authors of Ref. \cite{Du:2018gyn} seem to be aware of the problem since 
  they quote ``To investigate quantitatively possible poles beyond the near-threshold region, a more rigurous and complete treatement of the left hand cuts is required''. However, even then, they conclude the absence of the $f_2(1270)$ and $f'_2(1525)$ as dynamically generated resonances.
  
  \section{Brief description of the method of Ref. \cite{Geng:2016pmf}}\label{sectionII}
\begin{figure*}
  \centering
  \includegraphics[width=0.9\textwidth]{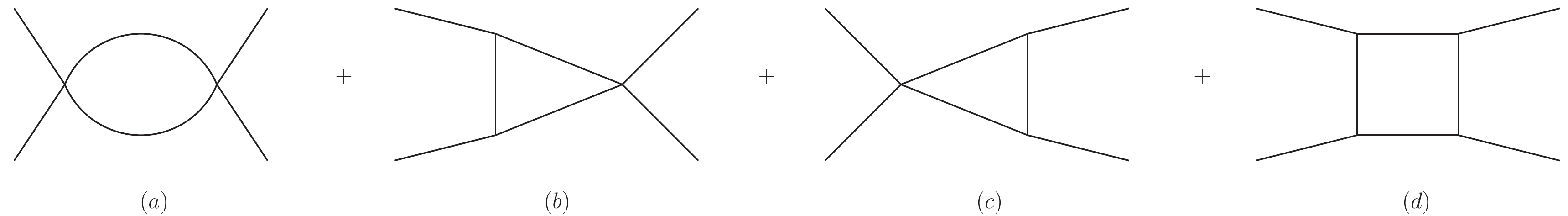} 
  \caption{Diagrams appearing at one-loop level with the contact and $\rho$-exchange terms.}\label{fig:2}
\end{figure*}

\begin{figure*}
  \centering
  \includegraphics[width=0.32\textwidth]{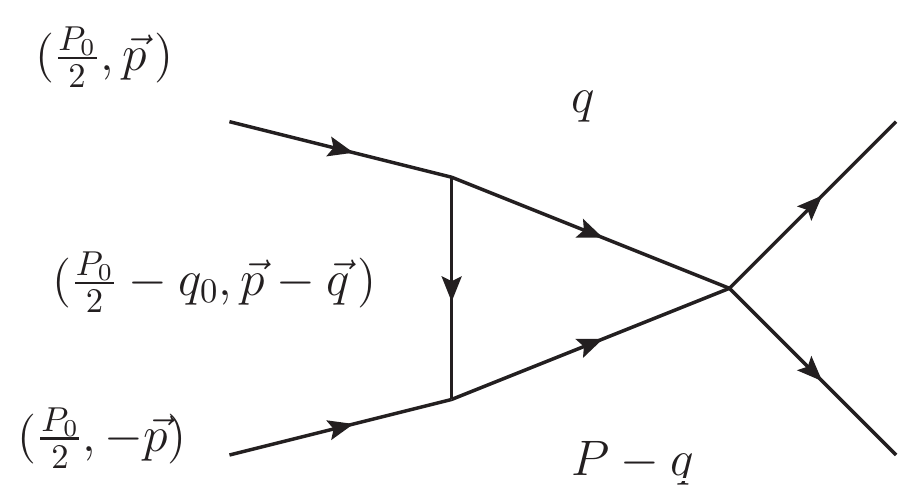} 
  \caption{Diagram of Fig.~\ref{fig:2}(b) showing explicitly the momenta of the particles.}\label{fig:3}
\end{figure*}

  In Ref. \cite{Gulmez:2016scm}, the propagator of the exchanged $\rho$-meson was projected in s-wave.
  \ba
  &&\hspace{-0.2cm}D_{\rho}(p)=\frac{1}{p^2-m^2_\rho+i\,\epsilon}\xrightarrow{s-wave} -\frac{1}{4\,p^2}\mathrm{Log}\left(\frac{4\,p^2+m^2_\rho}{m^2_\rho}+i\,\epsilon\right)\nonumber\\&&\hspace{-0.2cm} \equiv D_\rho^{(s.w.)}
 \label{eq:drho}\ea
 with $p^2=s/4-m^2_\rho$, on-shell, and (s.w.) denotes $s-wave$. This on-shell factorized propagator becomes infinite at $s=3\,m^2_\rho$ and for $s<3\,m^2_\rho$, $D_\rho^{(s.w.)}$ develops an imaginary part. In Ref. \cite{Geng:2016pmf} it was shown that the use of Eq. (\ref{eq:drho}), together with Eq. (\ref{eq:bethe}), leads to loop integrals below threshold which become infinite and have an imaginary part. 
 This evidences the deficiences of the method, since the one loop terms can be evaluated exactly, and so was done in Ref. \cite{Geng:2016pmf}. The results are finite and have no imaginary part below threshold.
 These loop diagrams are shown in Fig. \ref{fig:2}, and the momenta of diagram (b) are specified as in Fig. \ref{fig:3}. 
  The $t$-matrix for the diagram of Fig. \ref{fig:2} (b) after performing analytically the $q^0$ integral can be written as,
 \begin{align}
 t(s)=\int\frac{d^3q}{(2\pi)^3}\frac{1}{2\omega(q)^2}\frac{1}{2\,\omega(\vec{p}-\vec{q})}\frac{1}{P^0-2\,\omega(q)+i\,\epsilon}\nonumber\\\times\frac{1}{\frac{P^0}{2}-\omega(q)-\omega(p-q)+i\,\epsilon} t_c(-m^2_\rho)V_{\mathrm{ex}}V_c\ .
\label{eq:tt}\end{align}
In Eq. (\ref{eq:tt}) one has approximated the $\rho$ meson propagators with $q$ and $P-q$ momenta in Fig. \ref{fig:3} by their positive energy part, since they are placed close to on-shell in the loop, while for the exchanged $\rho$ with momentum $(P^0/2-q^0,\vec{P}-\vec{q})$, the full
$\rho$ propagator was kept. For the external $\vec{P}$ momentum, we take an average of the momentum for the wave function that the approach generates. Results are smoothly dependent on this value \cite{Geng:2016pmf}. Next one defines an effective $\rho$-meson exchanged propagator $D_{\rho,\mathrm{eff}}$
such that 
\be
(-m^2_\rho)V_\mathrm{ex} V_c D_{\rho,\mathrm{eff}}(s)G(s)=t(s)\ ,
\ee
with $G(s)$ the ordinary loop function which we regularize with the cut off method
\begin{equation}
\label{eq:loopco}
G=\int\limits_{|\vec{q}|\le q_\mathrm{max}}\frac{d^3q}{(2\pi)^3}\frac{\omega_1+\omega_2}{2\omega_1\omega_2[(P^{0})^2-(\omega_1+\omega_2)^2+i\epsilon]}
\end{equation}
where $q_\mathrm{max}$ stands for the cut-off, $(P^{0})^2=s$, and $\omega=\sqrt{\vec{q}^2+m^2_\rho}$. Then we define 
\be
\tilde{V}_{\mathrm{ex}}=(-m^2_\rho)V_{\mathrm{ex}}D_{\rho,\mathrm{eff}}
\ee
and by construction we have
\be
\tilde{V}_\mathrm{ex}V_cG(s)=t(s)\ .
\ee
With this $\tilde{V}_\mathrm{ex}$ we define the whole effective potential as
\be
V_{\mathrm{eff}}=V_c+\tilde{V}_\mathrm{ex}\label{eq:veff}
\ee
and if we do now
\be
V^2_{\mathrm{eff}}\,G(s)=(V_c+\tilde{V}_{\mathrm{ex}})^2G(s)\label{eq:veffg}
\ee
we are summing exactly the diagrams of Fig. \ref{fig:2} (a), (b) and (c), while it provides an approximation for the diagram of \ref{fig:2} (d). In Ref. \cite{Geng:2016pmf} the diagram of Fig. \ref{fig:2}(d) was also evaluated exactly and it was found that the approximation provided by Eq. (\ref{eq:veffg}), $\tilde{V}_\mathrm{ex} ^2G(s)$, differred from the exact result by $10$ \% around threshold and $18$ \% at $\sqrt{s}=1270$ MeV. Yet, taking into account the weight of all  
terms in Fig. \ref{fig:2}, Eq. (\ref{eq:veffg}), and the sum of the exact expressions for them, differred by $4.5$ \% at $\sqrt{s}=1270$ MeV and by $2.5$ \% at the $\rho\rho$ threshold. Then, $V_{\mathrm{eff}}$ was taken as an effective potential, and by means of 
\be
T=[1-V_{\mathrm{eff}}\,G]^{-1} V_{\mathrm{eff}}\ ,
\label{eq:goodt}
\ee
we could find poles for $J=0$ and $J=2$, similar to those found in Ref. \cite{raquel}.

In order to take into account the $\rho$ mass distribution in Ref. \cite{Geng:2016pmf} one has to take the function $G$ convoluted with the $\rho$ spectral function as

\begin{align}
\tilde{G}(s)= \frac{1}{N^2}\int^{(M_\rho+2\Gamma_\rho)^2}_{(M_\rho-2\Gamma_\rho)^2}d\tilde{m}^2_1(-\frac{1}{\pi}) {\cal I}m\frac{1}{\tilde{m}^2_1-M^2_\rho+i\Gamma\tilde{m}_1}\nonumber\\\times\int^{(M_\rho+2\Gamma_\rho)^2}_{(M_\rho-2\Gamma_\rho)^2}d\tilde{m}^2_2(-\frac{1}{\pi}) {\cal I}m\frac{1}{\tilde{m}^2_2-M^2_\rho+i\Gamma\tilde{m}_2} G(s,\tilde{m}^2_1,\tilde{m}^2_2)\ ,
\label{eq:Gconvolution}
\end{align}
with
\begin{equation}
N=\int^{(M_\rho+2\Gamma_\rho)^2}_{(M_\rho-2\Gamma_\rho)^2}d\tilde{m}^2_1(-\frac{1}{\pi}){\cal I}m\frac{1}{\tilde{m}^2_1-M^2_\rho+i\Gamma\tilde{m}_1}\ ,
\label{eq:orm}
\end{equation}
where $M_\rho=770$ MeV, $\Gamma_\rho=146.2$ MeV and for $\Gamma\equiv\Gamma(\tilde{m})$ we take the $\rho$ width for the decay into two pions in $p$-wave,
\begin{equation}
\Gamma(\tilde{m})=\Gamma_\rho (\frac{\tilde{m}^2-4m^2_\pi}{M^2_\rho-4m^2_\pi})^{3/2}\theta(\tilde{m}-2m_\pi).
\label{eq:gamma}
\end{equation}
Actually, it was found in Ref. \cite{Geng:2016pmf}, that if the $\rho$-meson exchange potential given by the propagator of Eq. (\ref{eq:drho}) is convoluted by the $\rho$-meson spectral function, it gives rise to a real part of the 
potential very similar to the one of Ref. \cite{raquel}. The infinity of the real part dissapears but the imaginary part remains although with no discontinuity. Taking into account the convolution of Eq. (\ref{eq:Gconvolution}) makes the problem more realistic, since now there are components of the $\rho\rho$ system which are actually not so bound even for the $f_2(1270)$.
\section{N/D approach of Ref. \cite{Du:2018gyn}}
Here we briefly comment on the N/D method used in Ref. \cite{Du:2018gyn}. In the approach, the scattering amplitude is given by
\be
T=N(s) D^{-1}(s)
\ee
with
\ba
&&N(s)=\sum_{m=0}^{n-1}\overline{a}'_m s^m+\frac{(s-s_0)^n}{\pi}\int^{s_{\mathrm{left}}}_{-\infty}d s'\frac{\mathrm{Im} T(s')D(s')}{(s'-s_0)^n(s'-s)}\ ,\nonumber\\
&&D(s)=\sum_{m=0}^{n-1}\overline{a}_m s^m+\frac{(s-s_0)^n}{\pi}\int^\infty_{s_{\mathrm{th}}} ds'\frac{\rho(s')N(s')}{(s'-s)(s'-s_0)^n}\ .\nonumber\\
\ea
Given the extreme difficulty of the exact solution, a perturbative approach is used in Ref. \cite{Du:2018gyn} approximating $N(s)$ by the potential $V(s)$, such that, for one channel, one has
\ba
&&N(s)=V(s)\ ;\nonumber\\
&&D_2(s)=\gamma_0+\gamma_1(s-s_{\mathrm{th}})+\frac{1}{2}\gamma_2(s-s_{\mathrm{th}})^2\nonumber\\&&+\frac{(s-s_\mathrm{th})\,s^2}{\pi}\int^{\infty}_{s_\mathrm{th}}ds'\frac{\rho(s')V(s')}{(s'-s_\mathrm{th}-i\,\epsilon)(s'-s-i\,\epsilon)s'^2}\ ,\nonumber \\ 
\label{eq:ds}\ea
with 
\ba
&&\rho(s)=\frac{\sigma(s)}{16\pi\,s}\ ;\qquad\sigma(s)=2\,p\sqrt{s}=\sqrt{(s-s_\mathrm{th})\,s}\ ,\nonumber\\\ea
where, in the last step, we have used that $m_1=m_2(=m_\rho)$ in the $\rho\rho$ channel, being $p$ the c.m. three momentum. 

The parameters $\gamma_0$, $\gamma_1$, $\gamma_2$ in Eq. (\ref{eq:ds}) are obtained matching $D_2(s)$ to $1-V\,G$ of Eq. (\ref{eq:bethe}) around the $\rho\rho$ threshold, or equivalently, matching
\ba\label{eq:subcs1}
&&P_2(s)\equiv\gamma_0+\gamma_1(s-s_{\mathrm{th}})+\frac{1}{2}\gamma_2(s-s_\mathrm{th})^2
\ea
to
\ba\label{eq:subcsnew}
&&\omega_2(s)=1-V(s)G(s)\nonumber\\
&&-\frac{(s-s_\mathrm{th}) s^2}{\pi}\int^{\infty}_{s_\mathrm{th}}ds' \frac{\rho(s')V(s')}{(s'-s_{\mathrm{th}}-i\,\epsilon)(s'-s-i\,\epsilon)s'^2}\nonumber\\
\ea
and then
\ba\label{eq:subcs}
&&\gamma_0=\omega_2(s_\mathrm{th});\qquad \gamma_1=\omega_2'(s_\mathrm{th});\nonumber\\
&&\qquad\gamma_2=\omega_2''(s_\mathrm{th})\ .
\ea

It is interesting to note that both $G(s)$  and the integral in Eq.~(\ref{eq:subcsnew}) have a discontinuity of the derivative at threshold. However the sum of the terms on the right hand side of  Eq.~(\ref{eq:subcsnew})  is well behaved as we show in the Appendix. Yet, a high accuracy in the numerical integrals is needed to accomplish it. We use Gauss integration with sufficient number of points to observe numerically the cancelation of the singular parts.

The final thing we want to show is that since $N(s)$ has been taken as $V(s)$, in the approach of Ref. \cite{Du:2018gyn}, where $D_\rho^{(s.w.)}$ of Eq. (\ref{eq:drho}) is used for the $\rho$-meson exchange potential, the $T$ matrix has unphysical singularities and 
imaginary part around $s=3\,m^2_\rho$. Hence, the method does not provide a realistic $t$-matrix. Yet, in Ref. \cite{Du:2018gyn}, the zeros of $D(s)$ are 
used to determine  whether the system has or not a pole, and $D(s)$ is a well behaved function since $V(s')$ is only used for $s'>s_\mathrm{th}$ and is never extrapolated to their unphysical region. 
Then it is interesting to see what happens.

  In order to understand the method and what it really accomplishes, we extend it to 
$(s-s_\mathrm{th})^3$ and compare the results with those at order $(s-s_\mathrm{th})^2$.

\section{Derivation of D(s) at ${\cal O}((s-s_\mathrm{th})^3)$}

In what follows we will compare the approximation of Eq.~(\ref{eq:ds}) for $D(s)$  to the exact value of $1-VG$. But before
that, we address the problem of extending Eq.~(\ref{eq:ds}) by making an extra subtraction and matching $1-VG$ up to $(s-s_\mathrm{th})^3$.

%

Eq.~(\ref{eq:ds}) contains one subtraction at threshold and two subtractions at $s=0$ to avoid problems with multiple subtractions at $s=s_\mathrm{th}$. Yet, the 
matching to $1-VG$ is done at threshold by means of Eqs.~(\ref{eq:subcs1}),~(\ref{eq:subcsnew}) and (\ref{eq:subcs}).~\footnote{One can rearrange a polynomial of order $s^3$ in $s$ in terms of a
polynomial in  $(s-s_\mathrm{th})$ up to order $(s-s_\mathrm{th})^ 3$.}  We make now one extra subtraction of the integral of Eq.~(\ref{eq:ds}) at $s=0$ and we obtain
\ba
&&D_3(s)=\gamma_0+\gamma_1(s-s_\mathrm{th})+\frac{1}{2}\gamma_2(s-s_\mathrm{th})^2+\frac{1}{3!}\gamma_3(s-s_\mathrm{th})^3\nonumber\\&&+\frac{(s-s_\mathrm{th})s^3}{\pi}\int^\infty_{s_\mathrm{th}}ds'\frac{\rho(s')V(s')}{(s'-s_\mathrm{th}-i\,\epsilon)(s'-s-i\,\epsilon)s'^3}\ .
\label{eq:ds3}\ea

To obtain the $\gamma_i$ coefficients we proceed as before and match
\ba
&&P_3(s)\equiv\gamma_0+\gamma_1(s-s_{\mathrm{th}})+\frac{1}{2}\gamma_2(s-s_\mathrm{th})^2+\frac{1}{3!}\gamma_3(s-s_\mathrm{th})^3\nonumber\\
\label{eq:subcs0}\ea
to
\ba\label{eq:subcsnew3}
&&\omega_3(s)=1-V(s)G(s)\nonumber\\
&&-\frac{(s-s_\mathrm{th}) s^3}{\pi}\int^{\infty}_{s_\mathrm{th}}ds' \frac{\rho(s')V(s')}{(s'-s_{\mathrm{th}}-i\,\epsilon)(s'-s-i\,\epsilon)s'^3}\nonumber\\
\ea
around the $\rho \rho$ threshold, and then
\ba\label{eq:subcs3}
&&\gamma_0=\omega_3(s_\mathrm{th});\qquad \gamma_1=\omega_3'(s_\mathrm{th});\nonumber\\
&&\qquad\gamma_2=\omega_3''(s_\mathrm{th});\qquad \gamma_3=\omega_3'''(s_\mathrm{th})\ .
\ea

\section{Wave function}
The wave function in momenta space reads \cite{YamagataSekihara:2010pj},
\be
\langle \vec{p}\vert \psi\rangle=A\frac{\Theta(p_\mathrm{max}-\vert\vec{p}\vert)}{E-\omega_1(p)-\omega_2(p)+i\epsilon}\ ,
\ee
where $\omega_{1,2}(p)=\sqrt{\vec{p}^2+m_{1,2}^2}$, and the normalization constant $A$ for a bound state can be obtained through the condition $A^2\int d^3 p\vert\langle p\vert \phi\rangle\vert^2=1$, as
\be
A=\sqrt{\frac{1}{\int_{p<p_\mathrm{max}} d^3 p \vert \frac{1}{E-\omega_1-\omega_2}\vert^2}}\ .\label{eq:norm}
\ee
While in coordinate space, throughout the Fourier Transform, we have,
\be
\langle \vec{r}\vert \psi\rangle=A\int_{p<p_\mathrm{max}}\frac{d^3p}{(2\pi)^{\frac{3}{2}}}e^{i\vec{p}\cdot\vec{r}}\frac{1}{E-\omega_1(p)-\omega_2(p)+i\epsilon}\ .
\ee
Since the exponential part can be decomposed in terms of the spherical Harmonic and Bessel functions,
\be
e^{i\vec{p}\cdot\vec{r}}=4\pi\sum_l i^l j_l(pr)\sum_m(-1)^mY_{lm}(\theta_{\hat{r}},\phi)Y_{l,-m}(\theta_{\hat{p}},\phi)\ ,
\ee
one can write down the wave function in coordinate space as,
\be
\langle \vec{r}\vert\psi\rangle=A\int_{p<p_\mathrm{max}}p^2dp\, 4\pi\, j_{0}(pr)\frac{1}{E-\omega_1(p)-\omega_2(p)+i\epsilon}\ ,
\ee
with $j_0(pr)=\frac{\mathrm{sin}(pr)}{pr}$. In the above relation, the condition $\int Y_{00}(\theta_{\hat{p}},\phi)Y_{0m}^*(\theta_{\hat{p}},\phi)d\Omega_{\hat{p}}=\delta_{m0}$ was used.
For the case of the $f_0(1370)$ and $f_2(1270)$, one needs to take into account the decay width of the $\rho$ meson. This can be done by convoluting the wave function with the $\rho$ meson mass distribution, like

\begin{align}
\widetilde{\langle \vec{r}\vert \psi\rangle}= \frac{1}{N^2}\int^{(M_\rho+2\Gamma_\rho)^2}_{(M_\rho-2\Gamma_\rho)^2}d\tilde{m}^2_1(-\frac{1}{\pi}) {\cal I}m\frac{1}{\tilde{m}^2_1-M^2_\rho+i\Gamma\tilde{m}_1}\nonumber\\\times\int^{(M_\rho+2\Gamma_\rho)^2}_{(M_\rho-2\Gamma_\rho)^2}d\tilde{m}^2_2(-\frac{1}{\pi}) {\cal I}m\frac{1}{\tilde{m}^2_2-M^2_\rho+i\Gamma\tilde{m}_2} \langle \vec{r}\vert \psi;\tilde{m}^2_1,\tilde{m}^2_2\rangle\ ,
\label{eq:Gconvolutionw}
\end{align}
with the normalization of Eq. (\ref{eq:orm}). For the case of open channels, as it occurs when taking into account the decay of the $\rho$ meson through the convolution of the wave function, where some $\rho\rho$ components are unbound, the wave function can become non normalizable, and we take the same normalization as in the bound case of Eq. (\ref{eq:norm}), which allows us to compare the wave function at small distances.

\section{Results}

\begin{figure}
  \centering
  \includegraphics[width=0.4\textwidth]{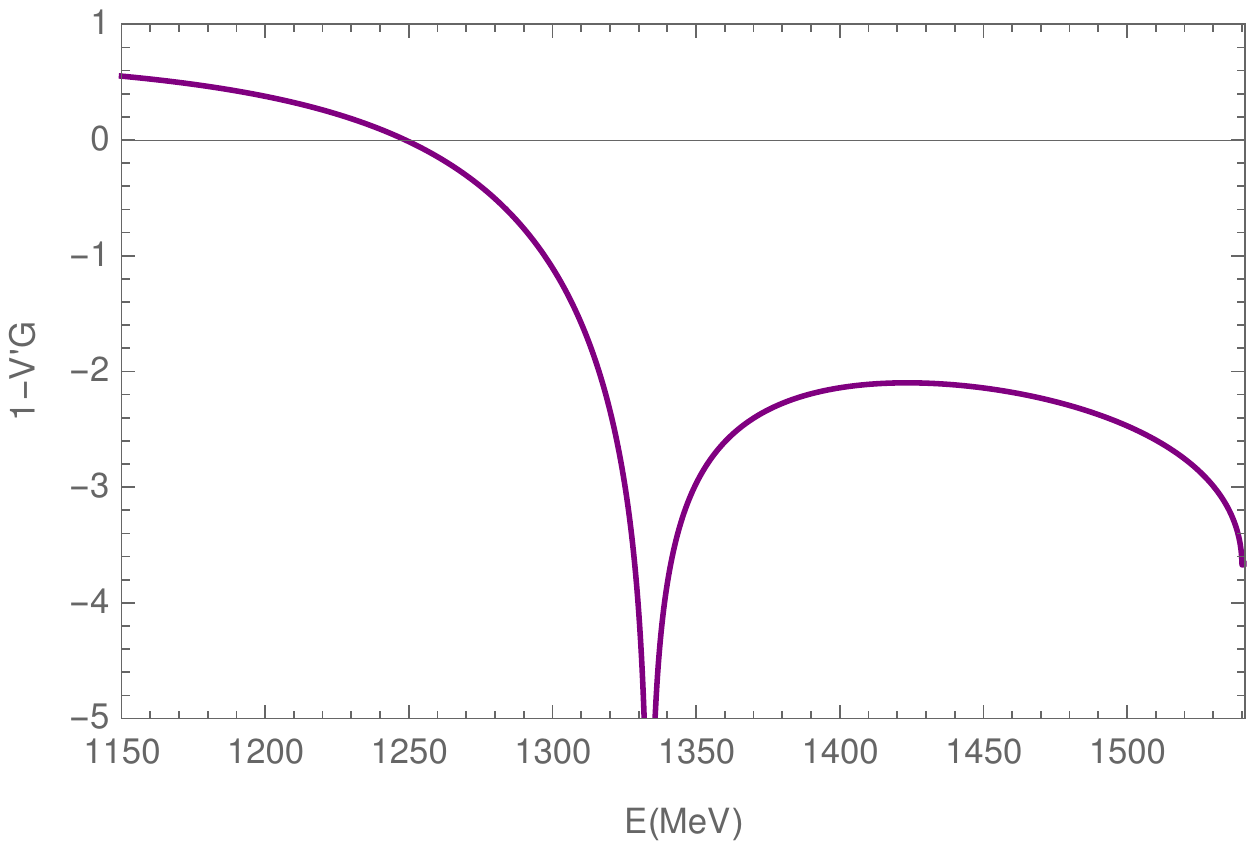} 
  \caption{The denominator of Eq. (\ref{eq:bethe}), $1-V'G$, with the potential of Eq. (\ref{eq:drhooller}), as a function of the energy.}\label{fig:new1}
\end{figure}

\begin{figure}
  \centering
  \includegraphics[width=0.45\textwidth]{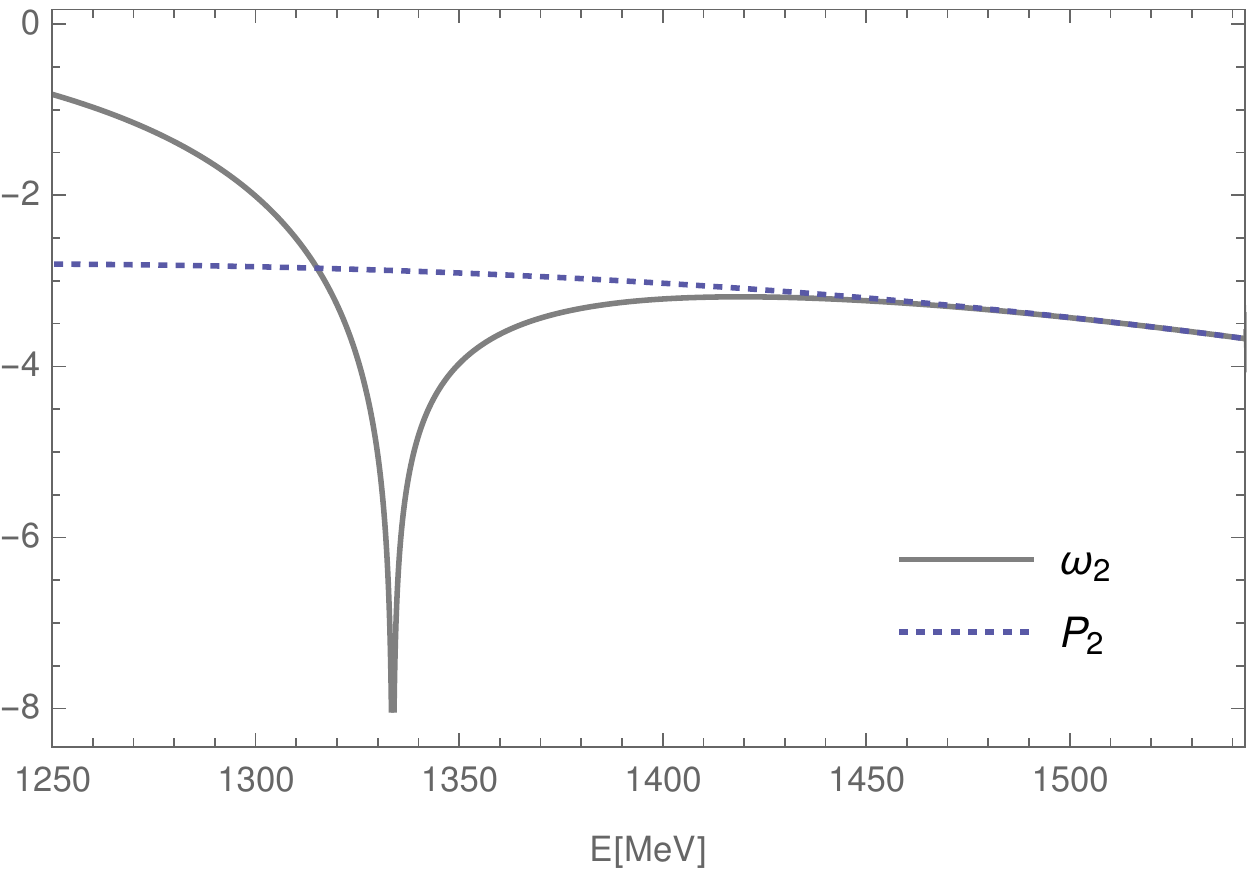} 
  \caption{The real part of the function $\omega_2$ in comparison with $P_2$, which appear in Eqs. (\ref{eq:subcs1}), (\ref{eq:subcsnew}), and (\ref{eq:subcs}),  with the potential $V'$.}\label{fig:new2}
\end{figure}

\begin{figure}
  \centering
  \includegraphics[width=0.45\textwidth]{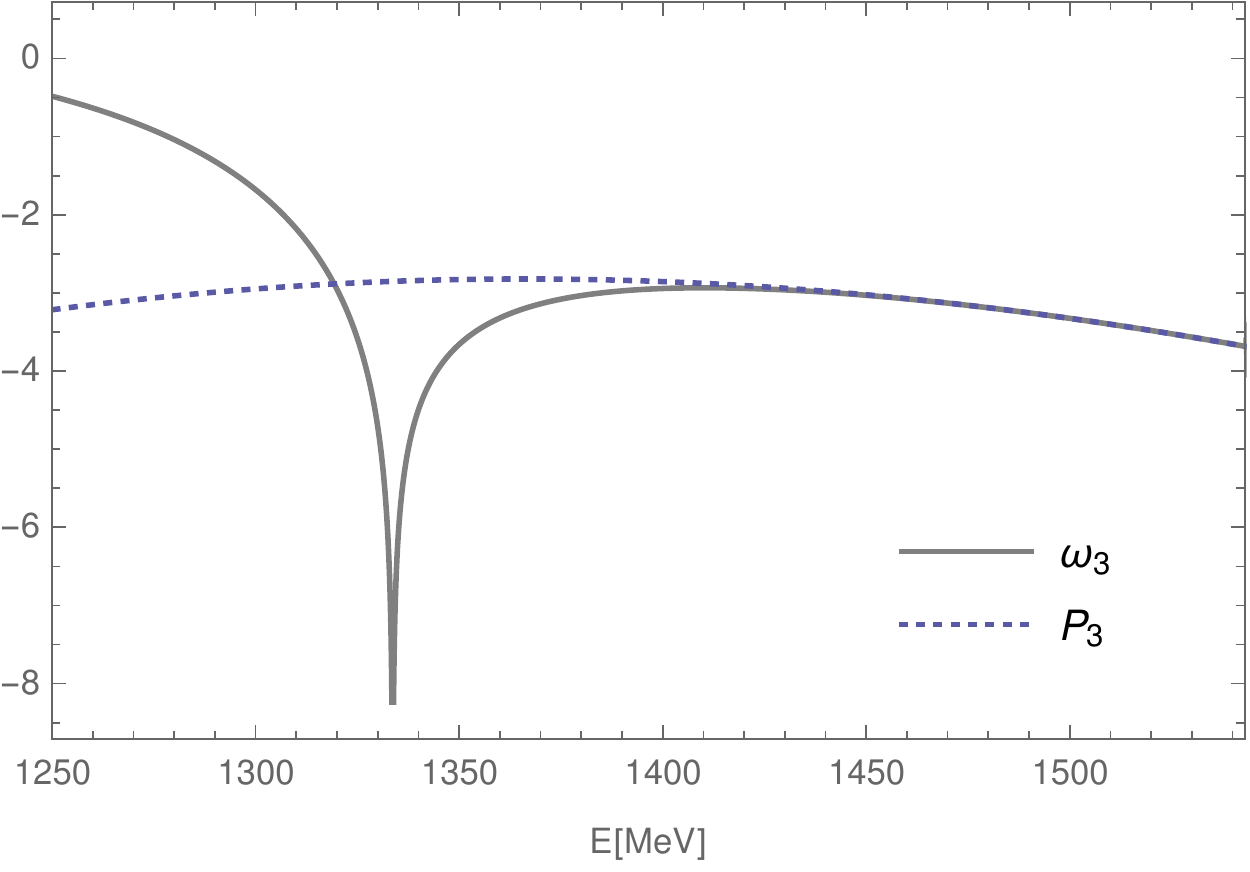} 
  \caption{The real part of the function $\omega_3$ in comparison with $P_3$ from Eqs. (\ref{eq:subcs0}), (\ref{eq:subcsnew3}), and (\ref{eq:subcs3}),  with the potential $V'$.}\label{fig:new3}
\end{figure}

\begin{figure}
  \centering
  \includegraphics[width=0.45\textwidth]{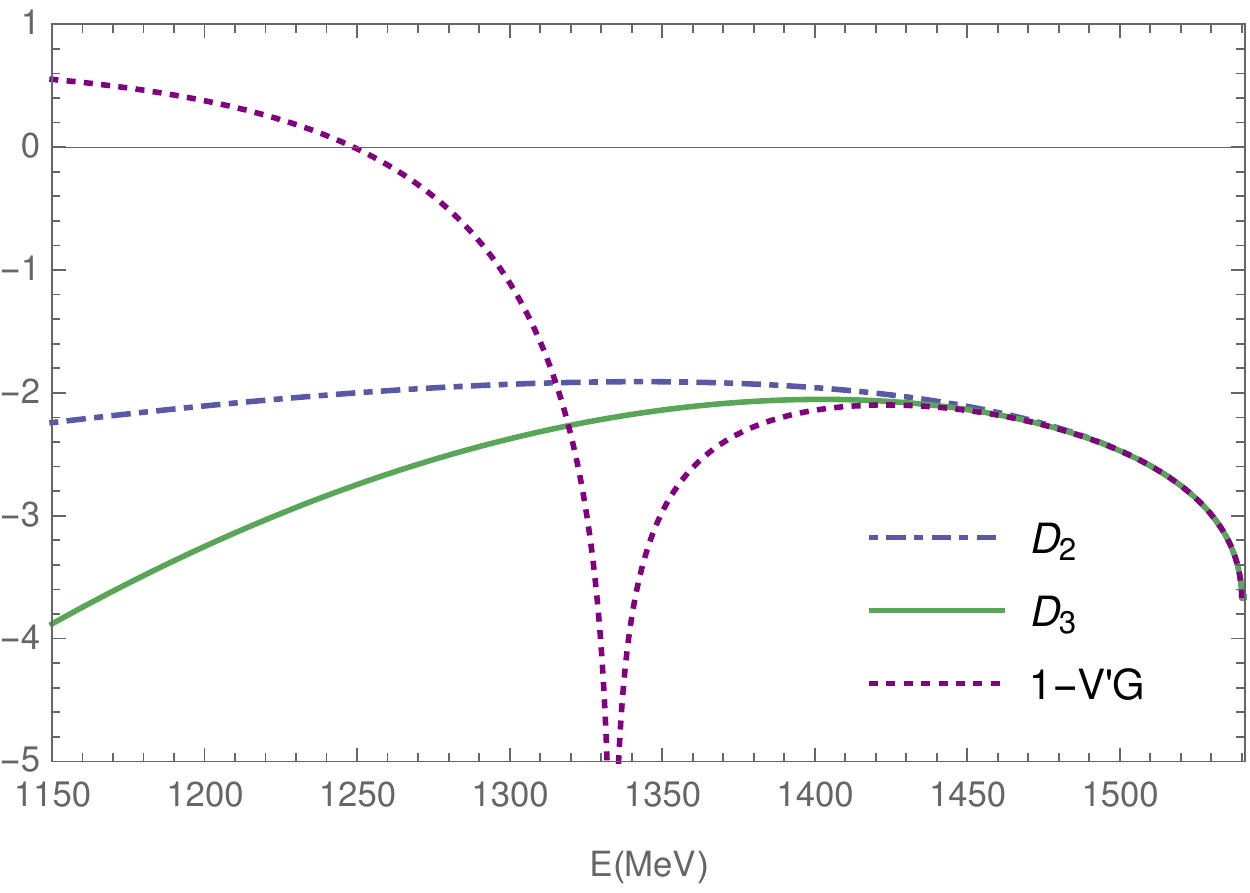} 
  \caption{The real part of the functions $D_2$ and $D_3$ from Eqs. (\ref{eq:ds}) and (\ref{eq:ds3}) in comparison with $1-V'G$. }\label{fig:new4}
\end{figure}

  Let us first study how the methods discussed previously work for the case of the singular potential, in which the projection over s-wave keeping the $q^2$ dependence of the $\rho$ propagator of Eq. (\ref{eq:drho}) is done, as used in Refs. \cite{Gulmez:2016scm} and  \cite{Du:2018gyn}. For this purpose, we take the sum of the contact term and the exchange term of Eq. (\ref{eq:ex}), but with $V_{ex}$ substituted by $V'_{ex}$, given by
\ba
&&V'_{ex}=V_{ex}(-m_{\rho}^2)D_\rho^{(s.w.)}\\
&&V'=V_c+V'_{\mathrm{ex}}\ .
\label{eq:drhooller}
\ea
In Ref. \cite{Gulmez:2016scm} extra terms were taken for the $3\rho$ vertex, which are negligible at energies close to threshold but are more relevant for lower energies. Yet, as noted in Ref. \cite{Geng:2016pmf}, the potential of Eq. (\ref{eq:drhooller}) is remarkably similar to the one of Ref. \cite{Gulmez:2016scm} shown in Fig. 4 of that work.
  In Fig. \ref{fig:new1} we plot the results for $1-V'G$ as a function of the total energy. We use $q_\mathrm{max}=1000$ MeV in $G$, Eq. (\ref{eq:loopco}), here and in the following \cite{raquel,Geng:2016pmf}. We see that a singularity appears around $E=1335$ MeV, corresponding to $s=3 m_\rho^2$. Let us see what we obtain using $D_2$ and $D_3$ from Eqs. (\ref{eq:ds}) and (\ref{eq:ds3}). This requires to evaluate first the functions $\omega_2(s)$ and $\omega_3(s)$ of Eqs. (\ref{eq:subcsnew}) and (\ref{eq:subcsnew3}). The parameters $\gamma_i$ which appear in $P_{2(3)}$, Eqs. (\ref{eq:subcs1}) and (\ref{eq:subcs0}), are obtained from a fit of $\omega_{2(3)}$ to these polynomials for energies around the threshold in a range of $5(10)$ MeV. In Figs. \ref{fig:new2} and \ref{fig:new3}, we plot $\omega_2$ together with the approximation by the quadratic polynomial $P_2$, and $\omega_3$ with the cubic approximation, $P_3$, respectively. The parameters $\gamma_i$ are shown in Table \ref{tab:ga2}.  We can see that both $\omega_2$ and $\omega_3$ are well behaved at threshold and are smooth functions of the energy, and that in both cases we obtain a  good fit to $\omega_2$ and $\omega_3$ by means of the polynomials $P_2$ and $P_3$ respectively, down to 1400 MeV.  Note also that $\omega_2$ and $\omega_3$ are not equal, since the integrals in their expression are not the same, and neither are the polynomials  $P_2$ and $P_3$ because two and three subtractions to the integrals, respectively, were done at $s=0$, instead of the threshold. However, when we evaluate $D_2$ and $D_3$, the two functions behave equally at threshold as a consequence of the fit that has been done to $1-V'G$. This can be seen in Fig. \ref{fig:new4}, where we plot $1-V'G$, $D_2$ and $D_3$. We can  see that, indeed, both  $D_2$ and $D_3$ are good approximations to $1-V'G$. The approximation with  $D_2$  is good down to $1450$ MeV, while with $D_3$ the approximation improves and is good down to $1400$  MeV.  However, at energies around 1270 MeV, where the $f_2(1270)$ resonance should appear, the two aproximatiosn differ appreciably from each other, although none of the two cuts the zero axis. This is essentially what is found in Ref. \cite{Du:2018gyn}, and from where it is concluded that the $f_2(1270)$ is not dynamically generated from the $\rho \rho$ interaction. However, the exercise of the expansion to order $(s-s_\mathrm{th})^3$ proves useful here. Indeed, what we see is that the approximation of $D_3$ tries to adjust better to $1-V'G$ in the upper part of the energies before the singular point appears. This cannot be otherwise, since the $D_2$, $D_3$ functions have been constructed precisely to avoid this singularity. There is no need to continue to higher orders in $(s-s_\mathrm{th})$ because one can see what would happen. Indeed, higher orders would bend more the curves around $1350$ MeV to adjust to $1-V'G$ in that region and would lead to a curve that would be below $D_3$. After many subtractions one would get close to the first branch of $1-V'G$ before the singularity. Certainly, there is no convergence of the different orders in the region below the singularity and hence, neither $D_2$ nor $D_3$, nor any higher order expansion, can be taken as a representation of a realistic $D$ function below the singular peak. Thus, the claim that the $f_2(1270)$ does not appear from the $\rho \rho$ interaction based on the approach of Ref. \cite{Du:2018gyn} is not justified. 

  After this exercise, let us perform another one that is illustrative. A minimum requirement when one deals with unstable particles is to perform a folding of the magnitudes with the spectral function (mass distribution) of these particles. Following this philosophy, we fold the $V'$ potential with the $\rho$ mass distribution using the same procedure as done to fold the $G$ function in Eq. (\ref{eq:Gconvolution}). This was done in Ref. \cite{Geng:2016pmf} and found to provide a real part very similar to the one of the potential used in Ref. \cite{raquel}. There is still one objection to use this potential since the convolution spreads the imaginary part that $V'$ artificially generates below threshold (see Fig. 5 of Ref. \cite{Geng:2016pmf}). Indeed, as discussed in detail in Ref. \cite{Geng:2016pmf}, the loops evaluated  using explicitly the full dynamics of the $\rho$ exchange do not have an imaginary part below threshold. This is because a) a bound state has a given energy and a wave function providing a distribution of real momenta, while the on shell factorization gives imaginary momenta. In the bound state the particles are not on shell. b) In the loops of the diagrams of Fig. \ref{fig:2} the two intermediate $\rho$ states in the s-channel can never be on shell if the external particles have an energy below threshold. As a consequence of that, and as was shown in Ref. \cite{Geng:2016pmf}, the exchanged $\rho$ mesons do not develop a singularity and the diagrams do not give any imaginary part. However, since the real part is similar to that of the potentials used in Refs. \cite{raquel} and \cite{Geng:2016pmf}, we perform the same exercise as before with this new potential, and the results are indicative of what one would get with the dispersion integral approach in all these other cases. The novelty of the convoluted potential is that the singularity disappears as soon as the convolution is done, as was shown in Ref. \cite{Geng:2016pmf}.
  
\begin{figure}
  \centering
  \includegraphics[width=0.45\textwidth]{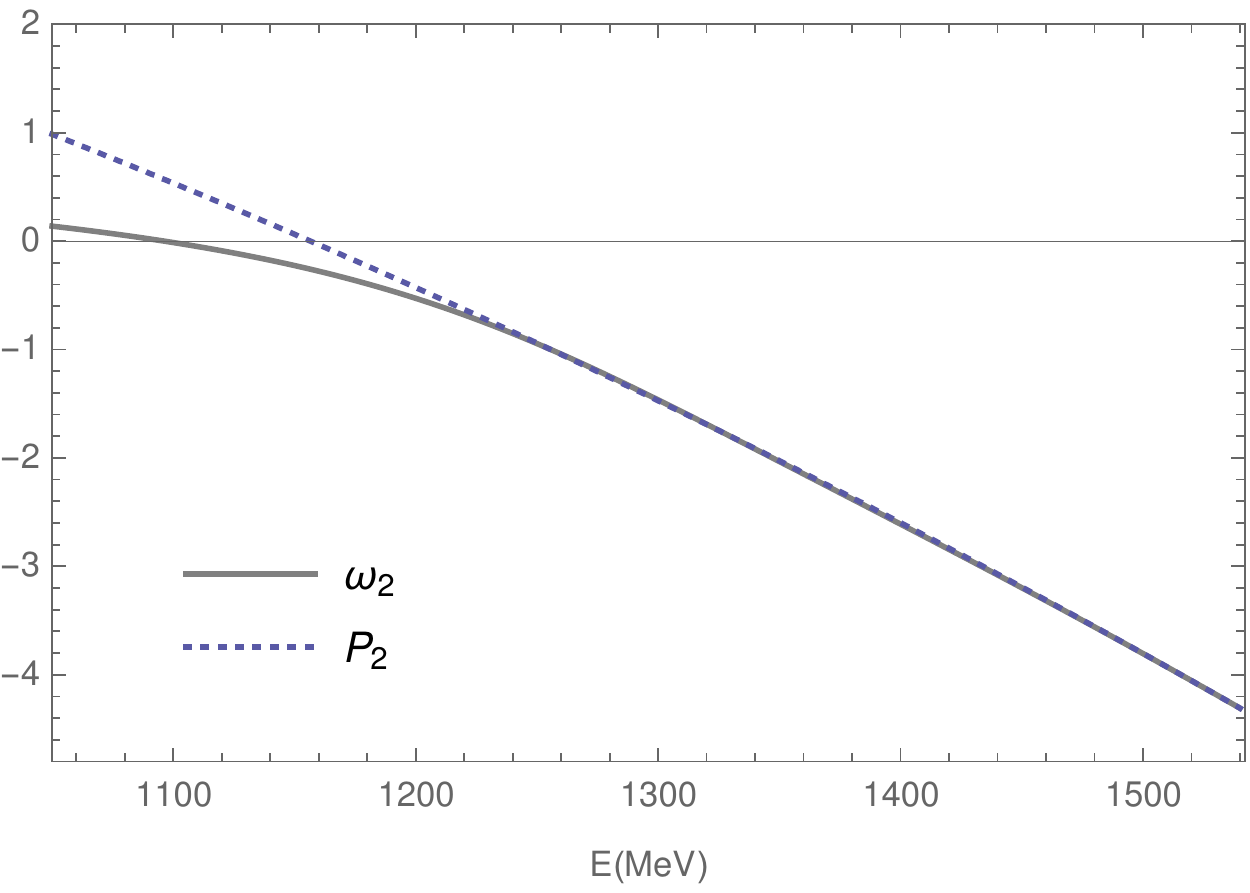} 
  \caption{The real part of the function $\omega_2$ in comparison with $P_2$, which appear in Eqs. (\ref{eq:subcs1}), (\ref{eq:subcsnew}), and (\ref{eq:subcs}),  with the convoluted potential $V$. }\label{fig:new5}
\end{figure}

\begin{figure}
  \centering
  \includegraphics[width=0.45\textwidth]{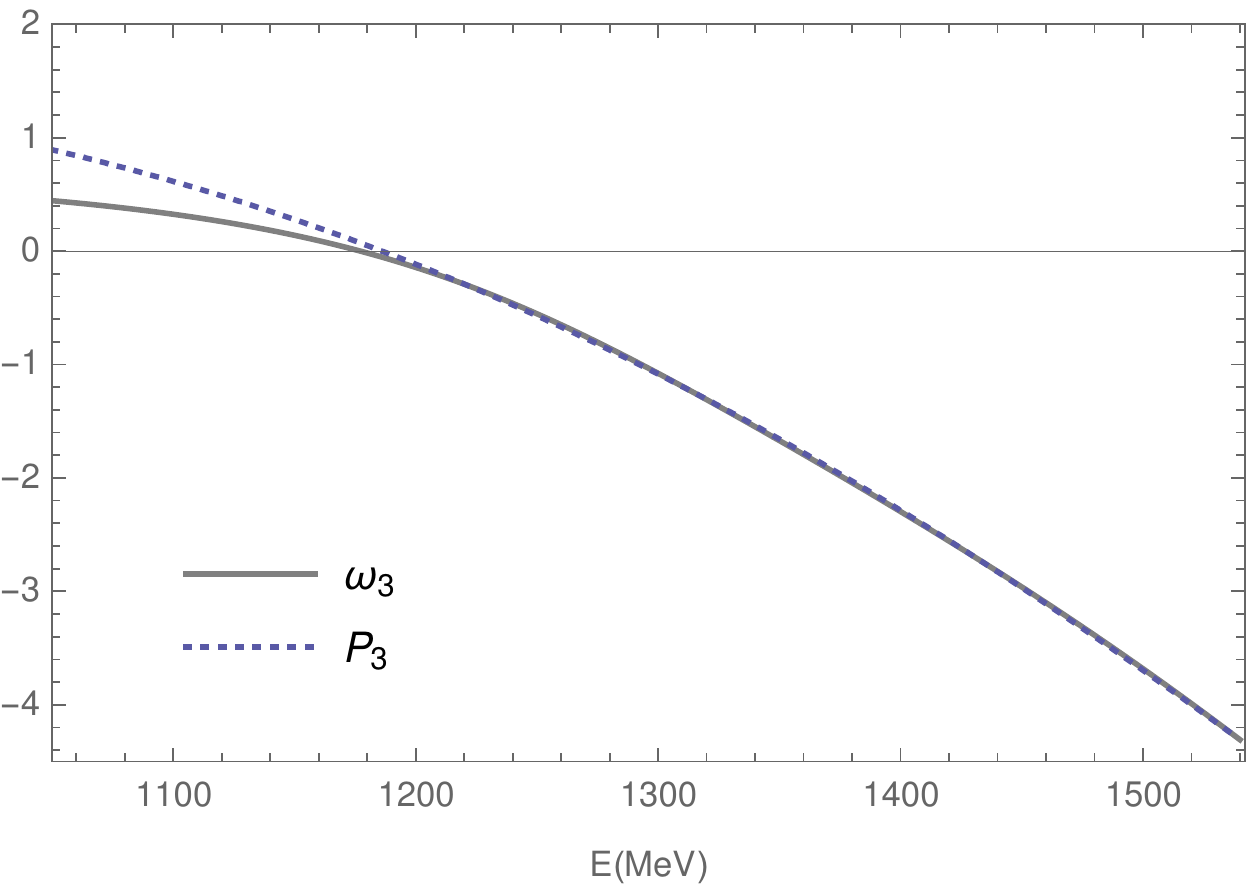} 
  \caption{The real part of the function $\omega_3$ in comparison with $P_3$ from Eqs. (\ref{eq:subcs0}), (\ref{eq:subcsnew3}), and (\ref{eq:subcs3}), with the convoluted potential $V$. }\label{fig:new6}
\end{figure}

\begin{figure}
  \centering
  \includegraphics[width=0.45\textwidth]{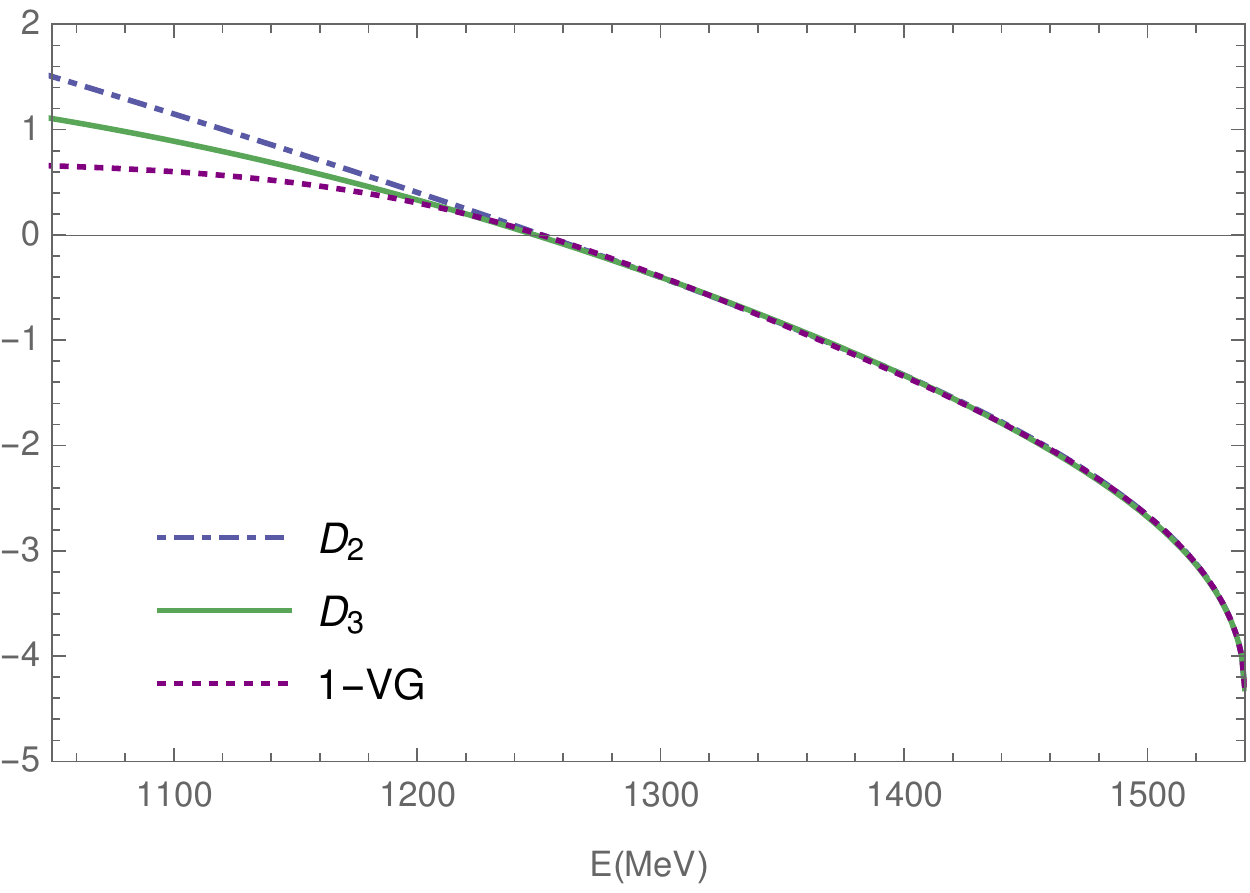} 
  \caption{The real part of the functions $D_2$ and $D_3$ from Eqs. (\ref{eq:ds}) and (\ref{eq:ds3}) in comparison with the real part of $1-VG$, with $V$ the convoluted potential. }\label{fig:new7}
\end{figure}

  In Figs. \ref{fig:new5} and  \ref{fig:new6}, we show again $\omega_2$ and $P_2$ for this new potential and $\omega_3$ together with $P_3$, respectively. The values of the $\gamma_i$ parameters are shown in Table \ref{tab:ga2}.  As happened before using the same potential without convolution, see Figs. \ref{fig:new2} and \ref{fig:new3}, 
$\omega_2$ and $\omega_3$ are well behaved below threshold, while  $P_2$ and $P_3$ are very good approximations to $\omega_2$ and $\omega_3$  respectively.   Next we plot $\mathrm{Re}\,D_2$ and $\mathrm{Re}\,D_3$, together with $\mathrm{Re}(1-VG)$, with $V$ the convoluted potential, and show the results in Fig. \ref{fig:new7}. We can see now that both $D_2$ and $D_3$ are good approximations to  $1-VG$ down to energies of $1200$ MeV. Moreover, in all these cases, the curves cut the zero axis around $1250$ MeV, the region where the $f_2(1270)$ appears.  This indicates that the dispersion approach provides a good convergence in a wide region of energies, provided the potential is not singular. However, in the case of the singular potential, we showed above that the approach provides unrealistic results for energies below the singular point and should not be used. 

    In the case of the realistic $V_{\mathrm{eff}}$ potential evaluated in Ref. \cite{Geng:2016pmf}, it was constructed such that the exact loops are generated by means of $V_{\mathrm{eff}}GV_{\mathrm{eff}}$, as discussed in section \ref{sectionII}, and hence Eq. (\ref{eq:goodt}) provides a realistic approach to the scattering matrix and generates a bound state around $1270$ MeV, as was shown in Ref. \cite{Geng:2016pmf}.


\begin{table*}[htb]
 \begin{center}
 {\renewcommand{\arraystretch}{2}
\setlength\tabcolsep{0.2cm}
\begin{tabular}{lrrrr}
\toprule
 Parameters: &$\gamma_0$&$\gamma_1\times 10^6\,(\mathrm{MeV}^{-2})$&$\gamma_2\times 10^{12}\,(\mathrm{MeV}^{-4})$&$\gamma_3\times 10^{18}\,(\mathrm{MeV}^{-6})$\\
 \hline
 $D_2$& $-3.7$ & $-2.0$ & $-2.4$ &- \\
 $D_3$& $-3.7$ & $-3.0$ & $-3.9$ & $7.7$\\
 \hline
 $D_2$& $-4.3$ & $-4.1$ & $0.04$ &- \\
 $D_3$ & $-4.3$&$-5.1$ &$-0.35$& $2.8$\\
\hline
 \end{tabular}}
\end{center}
\caption{Value of the parameters $\gamma's$ in Eqs. (\ref{eq:ds}), (\ref{eq:ds3}) using the on-shell potential of Eq. (\ref{eq:drhooller}) (upper two lines),  and with the convolution of Eq. (\ref{eq:Gconvolution}) (lower two lines).}
\label{tab:ga2}
\end{table*}

\begin{figure*}
\centering
  \hspace{-1cm}\includegraphics[width=0.9\textwidth]{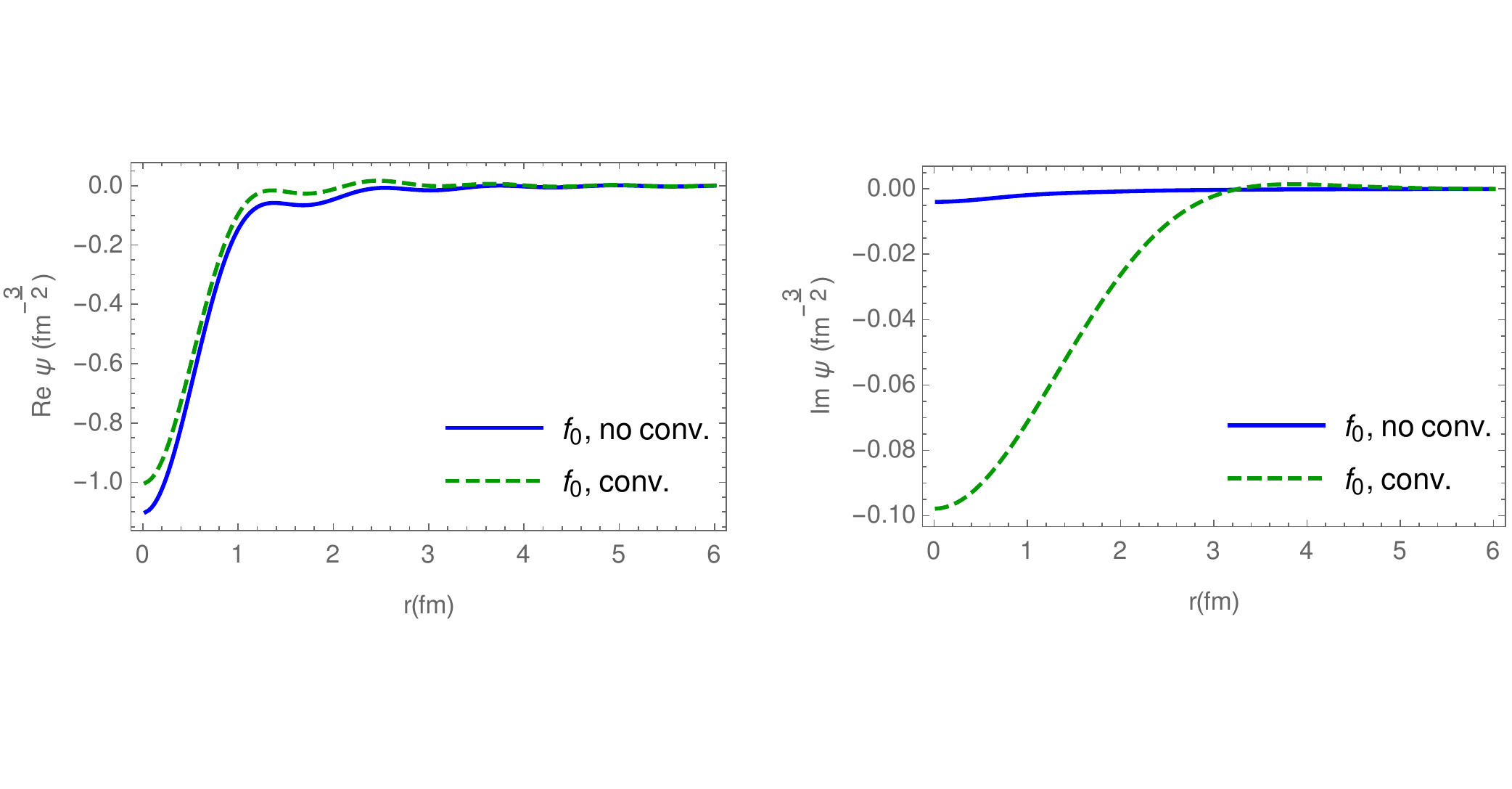}\vspace{-1.5cm}
  \caption{Real and imaginary part of the wave function in the coordinate space for the $f_0(1370)$, in the cases of no convolution and convolution with the $\rho$ meson spectral function.}\vspace{-1cm}
  \label{fig:wf0}
\end{figure*}

\begin{figure*}
\centering
  \hspace{-1cm}\includegraphics[width=0.9\textwidth]{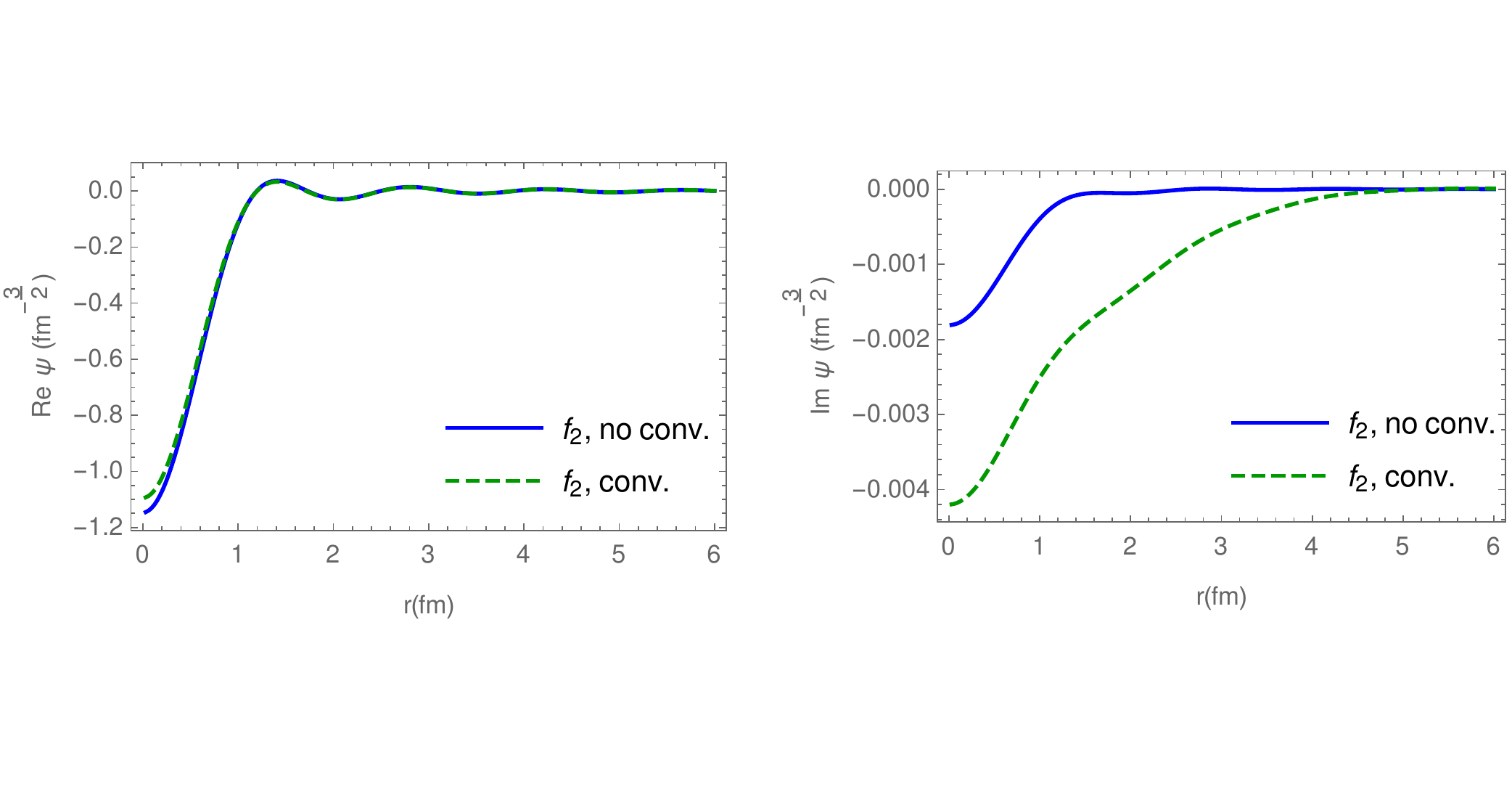}\vspace{-1.5cm}
  \caption{Real and imaginary part of the wave function in the coordinate space for the $f_2(1270)$, in the cases of no convolution and convolution with the $\rho$ meson spectral function.}
  \label{fig:wf2}
\end{figure*}

\begin{figure*}
{\renewcommand{\arraystretch}{2}
\setlength\tabcolsep{0.6cm}
\begin{tabular}{cc}
  \hspace{-0.8cm}\includegraphics[width=0.45\textwidth]{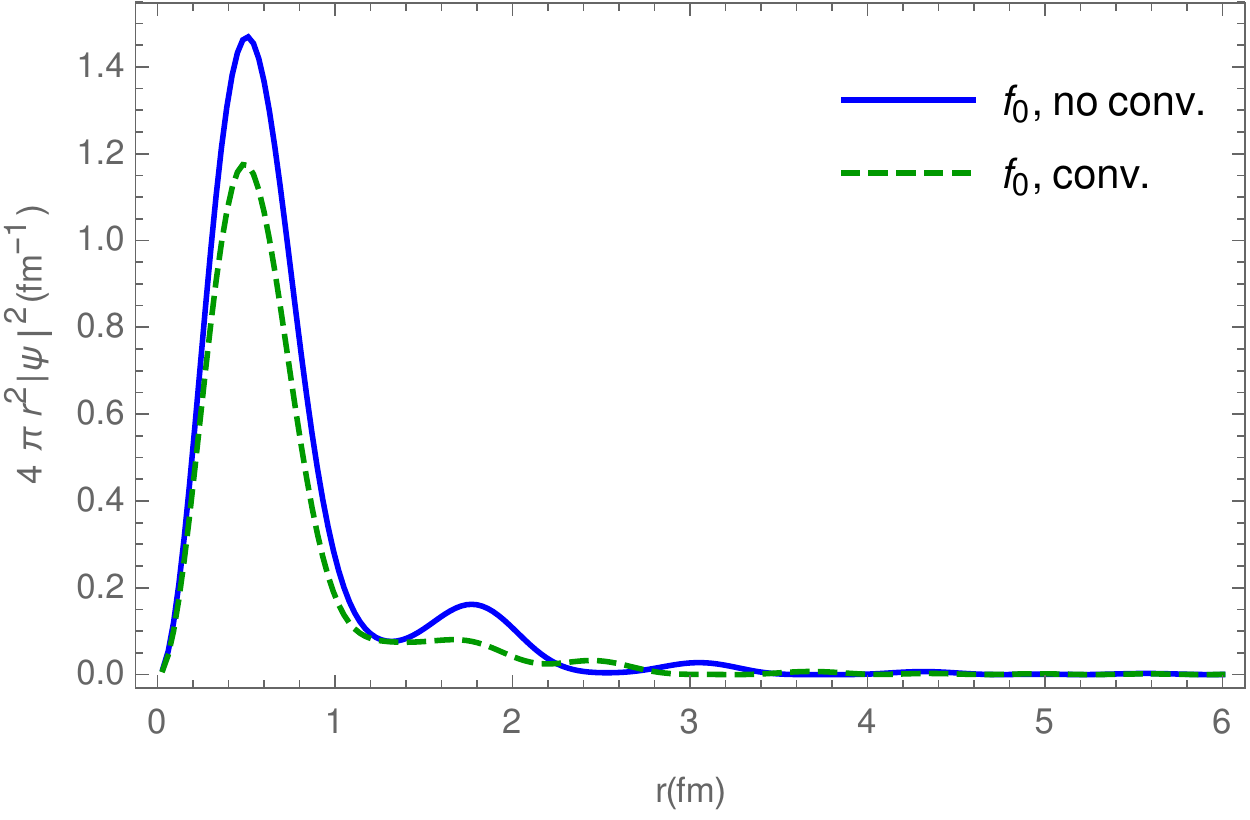}&\includegraphics[width=0.45\textwidth]{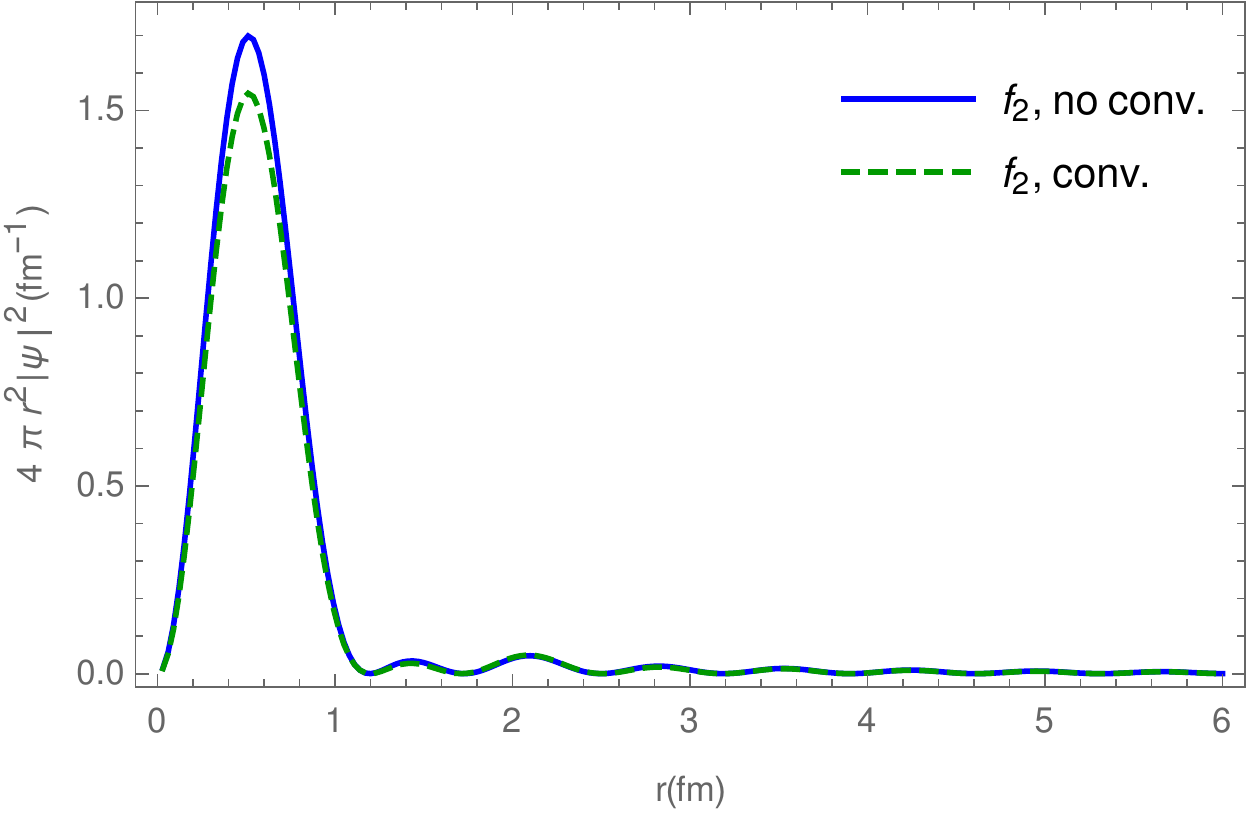}\\
  \end{tabular}}
  \caption{Probability density distribution in coordinate space for the $f_0(1370)$ (left) and $f_2(1270)$ (right), in the cases of no convolution and convolution with the $\rho$ meson spectral function.}
  \label{fig:awf02}
\end{figure*}

Finally, in Figs. \ref{fig:wf0} and \ref{fig:wf2}, we provide the result for the wave functions in coordinate space of the $f_0(1370)$ and $f_2(1270)$ in the cases of a completely bound state and a $\rho\rho$ resonance, when the $\rho$ meson is allowed to decay in two pions. We observe almost no difference if the convolution of the wave function is performed for the $f_2(1270)$, because of its larger binding energy. While for the $f_0(1370)$, the convolution has a bigger effect in the imaginary part of the wave function. For both resonances, the wave function for s-wave shows a peak at $r=0$. The probability density function, $4\pi r^2 \vert \psi\vert^2$, is depicted in Fig. \ref{fig:awf02}, peaking around $0.5$ fm. The oscillations in the wave function are caused by the sharp cut-off $p_\mathrm{max}$ around $1$ GeV used\footnote{We take $p_\mathrm{max}=1000$ MeV and $875$ MeV for the $f_0(1370)$ and $f_2(1270)$ respectively \cite{raquel}.}. Because of the larger binding energy, these are restricted in space till around $4-5$ fm, where the wave function approaches a value near zero.
\section{Conclusions}
We have analyzed in detail the method proposed in Ref. \cite{Du:2018gyn} to find out poles of the vector-vector scattering amplitudes, specializing to the $\rho\rho$ scattering. 
In order to avoid the use of an on-shell potential and the factorization in the Bethe-Salpeter equation proposed in Ref. \cite{Gulmez:2016scm}, since that potential develops unphysical singularities, the authors of Ref. \cite{Du:2018gyn} 
proposed an approach based on the $N/D$ method, performing some perturbative evaluation of $D(s)$, where the poles correspond to the zeros of $D(s)$. Apart from the approximation done to construct $D(s)$, an extra approximation is done to this $D(s)$  function performing
subtractions and fitting the results to $1-VG$ around threshold at order $(s-s_\mathrm{th})^2$. In the present work we extended that approach to order $(s-s_\mathrm{th})^3$ by making an extra subtraction to the dispersion integral. This allowed us to better understand what the dispersion approach is accomplishing. 

 What we found is that, in spite of the fact that the dispersion relation was introduced to avoid the unphysical divergence of the on shell $\rho$ exchange potential, the new $D$  function tries to adapt to $1-V'G$ with the singular potential in the region before the singularity and cannot be used to extrapolate to the region below this energy where the $f_2(1270)$ state appears. On the other hand we used a different potential, taking the same $\rho$ exchange term but folding it with the $\rho$ mass distribution. In this case the singularity disappears and the approach to 
$1-VG$ by means of the $D$ function of the dispersion relation is relatively good and can be extrapolated to relatively low energies. In this case we can see that $\mathrm{Re}\,(1-VG)$ and both $\mathrm{Re}\,D_2$ and $\mathrm{Re}\,D_3$ become zero at energies close to where the $f_2(1270)$ appears. However, $V$, $D_2$ and $D_3$, get an unphysical imaginary part. Yet, since $\mathrm{Re}\,V$ is very similar to the effective potential used in Ref. \cite{Geng:2016pmf}, the exercise done tell us what to expect in those realistic cases.

   In summary, the method proposed in Ref. \cite{Du:2018gyn} to avoid the pathologies of the use of the singular ``on-shell" $\rho$ exchange potential in Ref. \cite{Gulmez:2016scm}, eliminates indeed the artifical singularity of the  $D(s)$ function found in Ref. \cite{Gulmez:2016scm}, but we prove that its range of validity is constrained to energies much bigger than the one where the singularity appears and cannot be used to make predictions below this energy. But this is the case of the  $f_2(1270)$ resonance which appears below that point.

On the other hand, our approach with $V_\mathrm{eff}$, and $t=(1-V_\mathrm{eff}G)^{-1}V_\mathrm{eff}$, with $V_\mathrm{eff}$ constructed as in Eq.~(\ref{eq:veff}), gives a pole corresponding to a bound state in $J=2$,  the $f_2(1270)$. This state is unavoidable based on basic quantum Mechanics arguments since for $J=0$, where the potential has about one half the strength of $J=2$, all the methods produce a bound state, and hence for an attractive potential with double strength
a more bound state should be produced. 

\section*{Acknowledgments}
 We thank F. K. Guo for useful comments. This work is partly supported by the DGICYT contract FIS2011-28853-C02-01, FEDER funds from the European Union, 
 the Generalitat Valenciana in the program Prometeo, 2009/090, and 
the EU Integrated Infrastructure Initiative Hadron Physics 3
Project under Grant Agreement no. 283286. LSG acknowledges support from  
the National Natural Science Foundation of China under Grant  No. 11735003. R.M. acknowledges finantial support from the Fundac\~ao de amparo \`a pesquisa do estado de S\~ao Paulo, FAPESP (Ref. 2017/02534-3), and the Talento program from the  Community of Madrid (Ref. 2018-T1/TIC-11167).
\section*{Appendix: Cancelation of singularities in $\omega_2(s)$ and $\omega_3(s)$}
We make the derivation for $\omega_2(s)$. The case of $\omega_3(s)$ is identical. In the definition of $\omega_2(s)$, Eq. (\ref{eq:subcsnew}), $G(s)$ has a discontinuity in the derivative at threshold, and also the integral appearing there. Here we show that 
in the difference of the two terms these singularities cancel and $\omega_2(s)$ is well behaved at threshold. Let us begin with the integral
\begin{eqnarray}
 I_2(s)=\frac{(s-s_{\mathrm{th}})s^2}{\pi}\int^\infty_{s_{\mathrm{th}}}ds'\frac{\rho(s')V(s')}{(s'-s_{\mathrm{th}}-i\epsilon)(s'-s-i\epsilon)s'^2}\nonumber\\
\end{eqnarray}
with \begin{eqnarray}
\rho(s')=\frac{\sigma(s')}{16\pi s'};\qquad \sigma(s')=2p\sqrt{s'}=\sqrt{(s'-s_{\mathrm{th}})s'}\ ,\nonumber\\
     \end{eqnarray}
where $p$ is the momentum of one $\rho$ meson for the $\rho\rho$ system with energy $\sqrt{s'}$. It is convenient to work on the variable $p$,
\begin{eqnarray}
 &&2\sqrt{p^2+m_\rho^2}=\sqrt{s'};\qquad p^2+m^2_\rho=\frac{s'}{4};\qquad 2pdp=\frac{ds'}{4};\nonumber\\
&& p=\sqrt{\frac{s'}{4}-m^2_\rho}=\frac{1}{2}\sqrt{s'-s_\mathrm{th}};
\end{eqnarray}
Then, we have,
\begin{eqnarray}
 I_2(s)=\frac{(s-s_{th})s^2}{\pi}\int^\infty_0 \frac{dp\,p^2}{\pi \sqrt{s'}}\frac{V(s')}{(s'-s_{\mathrm{th}}-i\epsilon)(s'-s-i\epsilon)s'^2}\nonumber\\
\end{eqnarray}
Note that $p^2$ in the numerator cancels $(s'-s_\mathrm{th})$ in the denominator, showing that this denominator does not produce a singularity. Simplifying, we obtain,
\begin{eqnarray}
 &&I_2(s)=\frac{(s-s_{\mathrm{th}})s^2}{(2\pi)^2}\int^\infty_0 dp \frac{V(s')}{\sqrt{s'}(s'-s-i\epsilon)s'^2}\nonumber\\
 &&=\frac{(s-s_{\mathrm{th}})s^2}{(2\pi)^2}\int^\infty_0dp\frac{V(s')}{\sqrt{s'}(4p^2+4m_\rho^2-s-i\epsilon)s'^2}\nonumber\\
\label{eq:i2a}\end{eqnarray}
On the other hand, $G(s)$ from Eq. (\ref{eq:loopco}) gives 
\begin{eqnarray}
 G(s)=\int^{q_{\mathrm{max}}}_0 \frac{p^2 dp }{2\pi^2}\frac{1}{\omega(p)}\frac{1}{s-4 p^2-4m^2_\rho+i\epsilon}\label{eq:ga}
\end{eqnarray}
We can see that Eqs. (\ref{eq:i2a}) and (\ref{eq:ga}) have the same singular denominator. We can write $I_2(s)$ as
\begin{eqnarray}
 I_2(s)=-\frac{(s-s_{\mathrm{th}})s^2}{8\pi^2}\int^\infty_0  \frac{dp}{\omega(p)}\frac{V(s')}{(s-4p^2-4m^2_\rho+i\epsilon)s'^2}\nonumber\\
\end{eqnarray}
In order to see the singularity around threshold we can take $I_2(s)$ and $G(s)$ close to threshold and take $s^2/s'^2\equiv 1$, since 
the singularity comes from $s'=s_\mathrm{th}$. Thus, and just for values of $s$ very close to threshold and small values of $p$ in the integral,
\begin{eqnarray}
 &&1-VG(s)-I_2(s)\simeq \nonumber\\&&1-\frac{V(s_\mathrm{th})}{8\pi^2}\int\frac{dp}{\omega(p)}\frac{s-4m^2_\rho-4p^2}{s-4m^2_\rho-4p^2+i\epsilon}\ ,
\end{eqnarray}
and the singular denominator cancels with the numerator.

\end{document}